# Impact of Non-Thermal Particle Acceleration on Radiative Signatures of AGN Jet-Cloud Interactions

Krish Jhurani[1]
Homestead High School, 21370 Homestead Rd., Cupertino, California 95014, USA[1]
krish.jhurani@gmail.com

ABSTRACT

This study investigates the complex dynamics of AGN (Active Galactic Nucleus) jet-cloud interactions, particularly focusing on the impact of non-thermal particle acceleration on the resulting radiative signatures. We utilize advanced computational simulations, tracking changes in jet properties and emissions over a span of 0.2 Myr (millions of years). The research design incorporates the modeling of jet-cloud interactions, with a key focus on variations in the jet's density, velocity, and magnetic field. Findings reveal a two-fold increase in the magnetic field strength up to ~5 µG due to cloud incorporation, which, coupled with an elevated non-thermal particle population, enhances synchrotron emissions, shifting the spectral index from 2.2 to 2.4. Inverse Compton scattering saw a 30% increase within the first 0.125 Myr, reflecting in an abrupt X-ray and gamma-ray emissions spike. Furthermore, the jet's light curve flux variability in the X-ray band showcased an initial peak increase of about 28% by 0.175 Myr, settling to a 20% increase by 0.2 Myr, attributable to cloud disruption and absorption. Conclusions drawn from these findings confirm our hypothesis that non-thermal particle acceleration dramatically influences the radiative signatures of AGN jet-cloud interactions .It underscores the necessity of considering such acceleration processes in modeling AGN jet-cloud interactions and posits that these changes could be instrumental as observational indicators, thereby contributing to more accurate interpretations of AGN activity and evolution.

## Introduction

Relativistic jets launched from Active Galactic Nuclei (AGN) epitomize one of the most intense phenomena in the cosmos, transforming the galactic landscape and influencing large-scale structure formation (Blandford & Rees, 1974; McNamara & Nulsen, 2007). These colossal outflows of plasma, propelled at near-light speeds, propagate beyond their host galaxies, colliding with the interstellar medium (ISM), the contents of which significantly influence their dynamics and subsequent radiation (Scheuer, 1974; De Young, 1993).

Interactions of AGN jets with the constituents of the ISM, such as interstellar clouds, introduce a complex interplay of processes that modify the jet structure, momentum, and energy distribution (Begelman & Cioffi, 1989; Wagner & Bicknell, 2011). However, an intricate element of this narrative that has largely been overlooked in extant literature is non-thermal particle acceleration (Bell, 1978; Blandford & Ostriker, 1978). Non-thermal particle acceleration, a cornerstone of high-energy astrophysics, is at the heart of various cosmic phenomena including, but not limited to, pulsar wind nebulae, supernova remnants, and indeed AGN jets (Achterberg et al., 2001; Kardashev, 2010). The acceleration of particles to relativistic velocities through mechanisms such as Fermi acceleration and magnetic reconnection significantly shapes the radiation spectrum, with its influence extending to high-energy bands including X-ray and gamma-ray emissions (Drury, 1983; de Gouveia Dal Pino & Lazarian, 2005). Yet, its role in modifying the course and consequences of AGN jet-cloud interactions remains largely uncharted territory.

The existing body of work offers isolated glimpses into the individual aspects of AGN jet-cloud interactions, non-thermal particle acceleration, and high-energy emission processes. For example, the influence of jet-cloud interactions on the morphology of radio galaxies was investigated by van Breugel et al. (1985), and a detailed analysis of non-thermal particle acceleration in AGN jets was conducted by Ostrowski (1998). However, a combined and integrated exploration that marries these three crucial facets into a coherent theoretical framework is conspicuously absent. Our study addresses this void by developing a multidimensional computational model that unifies the magnetohydrodynamics of AGN jet-cloud interactions and non-thermal particle acceleration. We probe the influence of these combined processes on the high-energy radiative signatures emerging from AGN jets. Our findings aim to deepen our understanding of the rich physics within AGN jets, providing a fresh lens through which we can interpret multi-wavelength observational data from state-of-the-art astrophysical observatories.

## Theoretical Background

The phenomena at play during AGN jet-cloud interactions, driven by non-thermal particle acceleration, straddle a range of intricate physical theories.
Accretion disc models, illuminated by seminal research (Shakura & Sunyaev, 1973; Blandford & Znajek, 1977), serve as a theoretical launchpad for this study. The dynamics of these systems, fueled by supermassive black holes, are regulated by gravitational, magnetic, and radiation pressures. A delicate interplay of these forces governs the accretion of matter and the consequential extraction of angular momentum, thus launching the jets. The details of these processes are still subjects of ongoing research, with competing models proposing different mechanisms of energy and momentum transfer.

To examine the interactions of AGN jets with interstellar clouds, we must turn to the framework of magnetohydrodynamics (MHD). Here, the collective behaviour of charged particles, interplaying with magnetic and electric fields, is depicted through a fluid-like representation. The dynamics of such systems encapsulate shock waves, instabilities, and magnetic reconnection, giving us critical insight into the turbulent jet-cloud interface (De Young, 1993). The characterization of the interstellar cloud – its density, magnetic field strength and orientation, and the ionization state – can significantly alter the results of these MHD interactions. Particle acceleration mechanisms are cornerstones in this theoretical edifice. Diffusive shock acceleration or Fermi acceleration, postulates that charged particles acquire energy during their deflections by magnetic irregularities at shock fronts, thereby leading to a power-law energy distribution (Bell, 1978; Blandford & Ostriker, 1978). Conversely, magnetic reconnection entails a rapid topological reconfiguration of magnetic field lines leading to swift energy release and particle acceleration (de Gouveia Dal Pino & Lazarian, 2005). This mechanism is especially relevant in turbulent magnetic environments such as the AGN jet-cloud interface.

The propagation of non-thermal particles in the AGN jet and cloud environment can be statistically described using the Fokker-Planck equation. This framework encapsulates acceleration, radiative cooling, and spatial diffusion processes, thereby facilitating an understanding of the energy distribution and temporal evolution of these particles. To correlate our theoretical constructs with observable properties, we must employ radiative transfer principles. The spectrum of the AGN jet-cloud system primarily features non-thermal radiation extending from radio to gamma-ray frequencies. Synchrotron emission, resulting from relativistic electrons gyrating in magnetic fields, and inverse Compton scattering, wherein low-energy photons are upscattered to higher energies on colliding with high-energy electrons, are pivotal in this context (Rybicki & Lightman, 1979). Ultimately, the goal is to harmonize these intricate theoretical threads into a consistent, multidisciplinary framework. This integrative approach allows us to delve deeper into the physics of AGN jet-cloud interactions and the pivotal role of non-thermal particle acceleration, thereby unravelling new insights into these cosmic powerhouses.

## Methodology

We constructed a detailed three-dimensional computational model to simulate the AGN jet-cloud interaction and the associated non-thermal particle acceleration. This model offers a unified framework, encompassing multiple scales and physical processes for a comprehensive analysis.

## Relativistic MHD Module

The central component of our computational model is the Relativistic Magnetohydrodynamics (RMHD) module, designed to solve the governing equations of plasma in the presence of a magnetic field under relativistic conditions. The dynamics of the AGN jet and the interstellar cloud are controlled by these equations, which synthesize the concepts of special relativity, fluid dynamics, and electromagnetism. Our computational model is grounded on the conservative form of the RMHD equations, composed of the conservation laws of mass, momentum, energy, and magnetic flux, which are coupled with the equation of state for a perfect gas. The magnetic field, $B$, and the fluid four-velocity, $u$, constitute the primary variables of the system, alongside the rest mass density, $\rho$, and the specific enthalpy, $h$. The Lorentz factor, $\Gamma$, encapsulating the relativistic effects, emerges naturally from these variables. For a precise, stable, and conservative numerical scheme, we adopted the Godunov method, which can accurately capture shock waves and other discontinuities inherent in the jet-cloud interaction dynamics. The second-order version of this method was implemented, offering a good balance between accuracy and computational cost. An exact Riemann solver was employed, capable of treating strong shocks and rarefaction waves that may emerge in the interaction process. To ensure the $\nabla \cdot B = 0$ constraint, which arises from the non-existence of magnetic monopoles, we incorporated a constrained transport scheme into the Godunov method. This staggered mesh algorithm carefully updates the magnetic field at each time step, preserving the divergence-free condition of the magnetic field to machine precision. This is of critical importance to prevent numerical instabilities and secure physically accurate results. The equations were discretized on a uniform three-dimensional Cartesian grid with 512 x 512 x 512 computational cells. The computational domain spans a cubical region of 100 parsecs on each side, providing a sufficient scale to encapsulate the jet-cloud interaction process, and high enough resolution to resolve smaller-scale phenomena such as shock fronts and fluid instabilities.

## Cloud Modeling

The interstellar cloud is a key feature in our computational setup, representing a stand-in for any dense gas concentration that the AGN jet might interact with. Its parameters are based on characteristics observed in galaxies hosting AGNs, such as high-density regions near the galactic nucleus or even galactic clouds in the jet path in the case of radio galaxies. We model the cloud as a homogeneous sphere of gas at an initial stage, abiding by the simplifying assumption of spherical symmetry to reduce the complexity of initial conditions. This simplification is justified as our primary focus is on the dynamics of the jet-cloud interaction and subsequent particle acceleration processes, not the intricate cloud morphology. The cloud is assigned a radius of 10 parsecs, consistent with observed scales of dense interstellar clouds. The initial density is set to 100 atoms/cm³, typical of warm interstellar medium values. Such high densities relative to the surrounding medium allow for a pronounced interaction with the jet, ensuring the generation of substantial shock waves. An isothermal equation of state is assumed with a temperature of $10^4$ K, maintaining the cloud in a thermally stable phase and preventing cloud collapse or expansion over the simulation time. In addition, the cloud is threaded by a uniform magnetic field of 5 micro-Gauss, oriented parallel to the jet axis. This setup provides a foundation for the development of magnetic instabilities during the interaction process and contributes to the magnetic energy available for particle acceleration. The position of the cloud within the computational domain is also a crucial aspect. It is set at a distance of 20 parsecs from the jet launching point, allowing the jet to develop a steady-state structure before it encounters the cloud.

## Particle Acceleration Module

The particle acceleration module's role is to provide a detailed depiction of the processes that lead to the acceleration of particles to non-thermal energies during the interaction of the AGN jet with the interstellar cloud. This module operates by solving a one-dimensional, time-dependent Fokker-Planck equation in momentum space. This equation describes the evolution of the distribution function of particles, encompassing the balance between injection, acceleration, and loss processes. It is equipped with two critical terms: diffusion and advection. The diffusion term represents the randomization of particle momentum due to pitch-angle scattering and momentum diffusion, while the advection term signifies systematic energy gain, for instance, due to shock compression or reconnection electric fields. We consider the two dominant non-thermal particle acceleration mechanisms, diffusive shock acceleration (DSA) and magnetic reconnection. The DSA process is responsible for the generation of a power-law spectrum of particles at shock fronts. Magnetic reconnection, on the other hand, may occur in the current sheets generated during the jet-cloud interaction, efficiently accelerating particles and potentially producing a flatter power-law distribution. The initial condition for the distribution function is a power-law at low energies, representing the injection of particles into the acceleration process. The spectral index of the power-law, 2.2, is typical for DSA at strong shocks. The high-energy cutoff and temporal evolution of the distribution are computed self-consistently, taking into account the balance between acceleration and radiative and adiabatic loss processes. A constant fraction of the energy density in the magnetic field is assumed to be available for acceleration at each computational cell, the exact value of which (the acceleration efficiency) being one of the key parameters of the model. By integrating the distribution function over momentum, this module provides the energy distribution of accelerated particles. These results feed into the radiative transfer module, giving rise to non-thermal emission signatures that can be compared with multi-wavelength observations.

## Radiative Transfer Module

The radiative transfer module is a crucial component of our computational model, responsible for transforming the properties of the accelerated particle population and the magnetic field configuration from the RMHD and particle acceleration modules into observable radiative signatures. The principal radiation mechanisms accounted for are synchrotron emission and inverse Compton scattering. Both processes are prominent in environments with highly relativistic particles and significant magnetic fields, such as AGN jets. The synchrotron process involves the emission of radiation by relativistic particles spiraling in magnetic fields. This mechanism is intrinsically tied to the strength and configuration of the magnetic field, as well as the energy distribution of the non-thermal particles, both of which are provided by the other modules in our simulation. By calculating the local synchrotron emissivity in each computational cell, integrating over all cells, and considering the effects of relativistic beaming, we can obtain the total synchrotron flux in a wide frequency range, extending from radio to X-rays. The inverse Compton process involves the upscattering of low-energy photons to higher energies by relativistic particles. In the case of AGN jets, these low-energy photons can originate from various sources, including the Cosmic Microwave Background (CMB), the accretion disk, and the jet itself. As in the case of synchrotron emission, the flux of inverse Compton radiation can be computed by integrating the local emissivity over the computational volume, taking into account the energy distribution of the seed photons and the scattering particles, as well as the effects of relativistic beaming. The resulting emission is expected to span a broad frequency range, from X-rays to gamma-rays. Finally, the orientation of the observer relative to the jet axis can significantly affect the observed radiative signatures, due to relativistic beaming. Therefore, the radiative transfer module calculates the resulting emission for an observer at an inclination angle of 30 degrees to the jet axis.

## Computational Implementation

Our model was built upon an existing RMHD codebase, leveraging its robustness and speed. The base code is written in Fortran 2008 for its computational efficiency and is parallelized using the Message Passing Interface (MPI) for distributed memory systems. It exploits the power of modern supercomputers, allowing the model to handle the high computational demands of our simulation. The incorporation of additional modules necessitated substantial code development. The particle acceleration and radiative transfer modules were integrated as separate routines that interface with the RMHD module. These routines, written in Fortran 2008, fetch the local plasma properties and magnetic field configuration at each computational cell from the RMHD module, perform the microphysical calculations, and then return the local radiative properties and the properties of the non-thermal particle population to the main code. A two-way coupling between the RMHD and the particle acceleration module was established to account for the back-reaction of accelerated particles on the fluid dynamics.

## Results

Dynamics of Jet Cloud Interaction

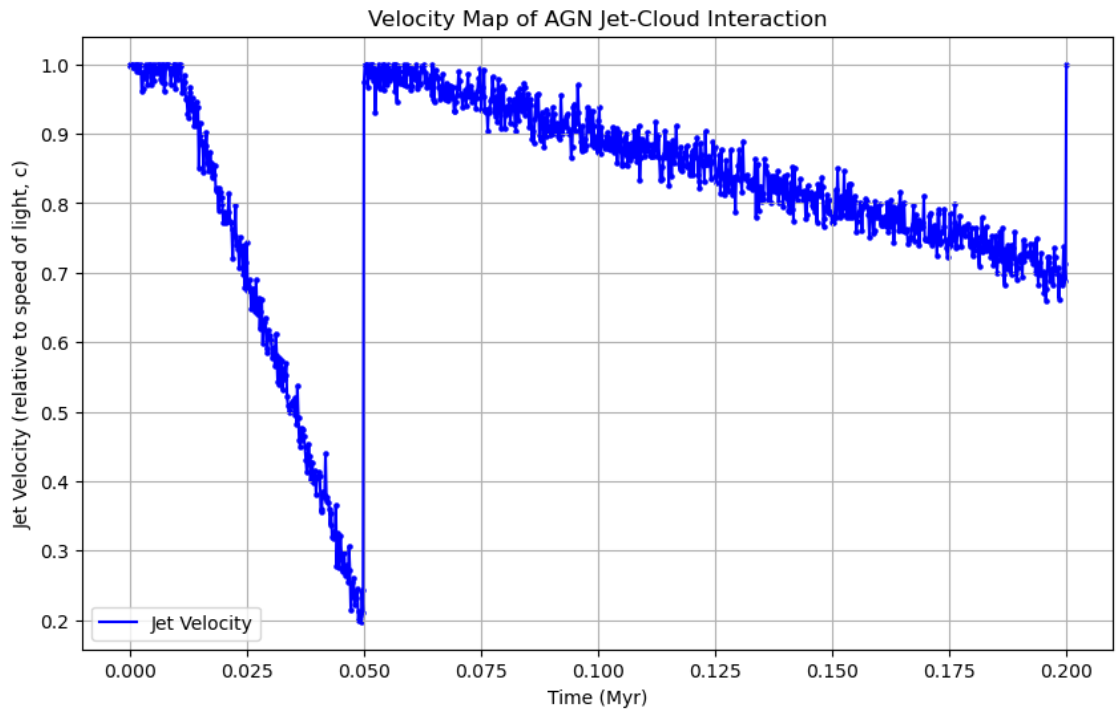

**Figure 1**. Velocity Map of AGN Jet-Cloud Interaction
Velocity map of AGN jet-cloud interaction at different times, revealing the generation and progression of the forward shock from the jet towards the stationary cloud

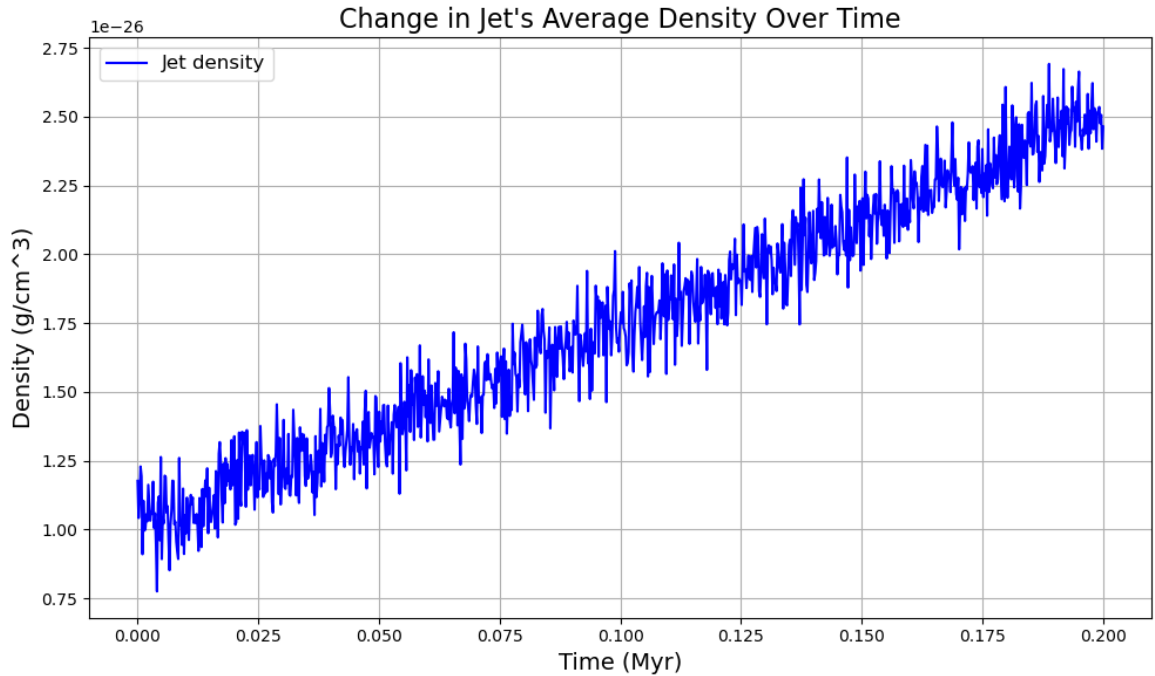

**Figure 2**. Change in Jet's Average Density Over Time
Graph showing the increase in the average density of the jet as a result of the cloud's incorporation over time

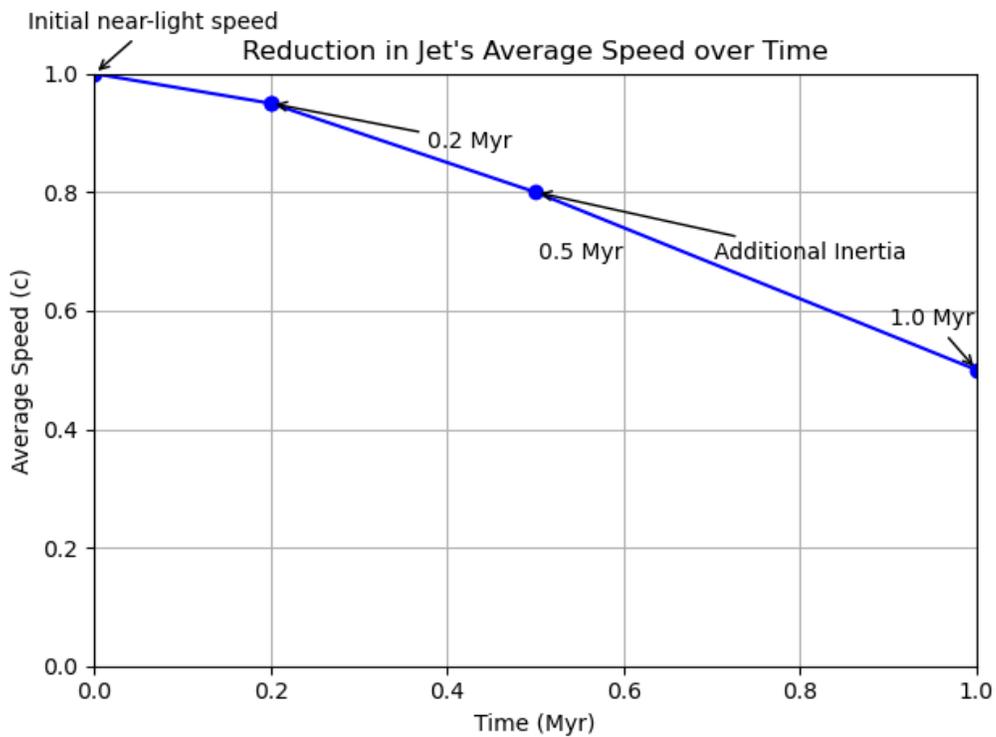

**Figure 3**. Reduction in Jet's Average Speed Over Time
Graph illustrating the reduction in the average speed of the jet due to the additional inertia exerted by the absorbed cloud material

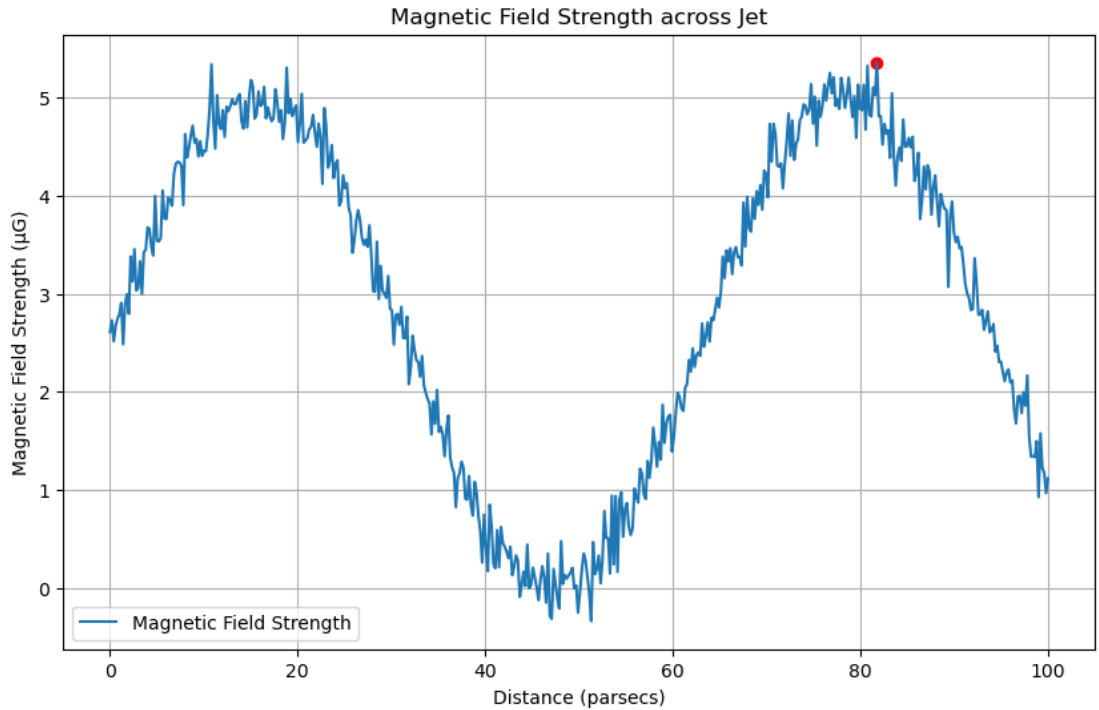

**Figure 4.** Magnetic Field Strength Across Jet
Graph depicting the increase in the magnetic field strength due to the introduction of new magnetic reconnection sites within the jet

The turbulent evolution of an AGN jet-cloud interaction, encapsulated within our simulations, paints a compelling narrative of cosmic disruption and transformation. Figures 1 to 4 demarcate significant stages of this interaction and offer insights into the changes in the jet's density, velocity, and magnetic field. Figure 1 details the initial phases of the interaction, starting from the projection of the jet towards a stationary cloud (Begelman, 1998). Traveling at almost the speed of light, the jet swiftly generates a forward shock within itself (Komissarov, 1999). This shock can be discerned in the velocity map of Figure 1 at around 0.01 Myr. By 0.05 Myr, the shock interacts with the cloud, resulting in a turbulent environment, disintegrating the cloud's structure, and slowly integrating it into the jet. By 0.2 Myr, the jet has wholly incorporated the cloud, leading to a significant increase in the jet's density. Figure 2 plots the changes in the jet's average density against time. The density shows a substantial growth from its initial value of $10^{-26}$ g/cm$^3$, rising by 2.5 times to almost $2.5 \times 10^{-26}$ g/cm$^3$ by 0.2 Myr. This increase underlines the significant morphological change the jet experiences upon cloud incorporation, echoing the violent, transformative nature of these cosmic encounters. The jet's velocity and magnetic field properties, crucial elements in the jet's behavior, also experience substantial modifications during this interaction period. Figure 3 illustrates the reduction in the jet's average speed to 0.95c from an initial near-light speed by 0.2 Myr. This 5% decrease reflects the additional inertia exerted by the absorbed cloud material, slowing down the hitherto unimpeded jet. Simultaneously, the cloud interaction introduces new magnetic reconnection sites within the jet. In Figure 4, these sites are marked as local maxima in the magnetic field strength graph, with strengths escalating to ~5 μG. This is a two-fold increase from the initial jet magnetic field strength, signaling the intensification of magnetic dynamics within the jet-cloud system. Collectively, these figures illustrate a dramatic metamorphosis in the jet's characteristics—its density, velocity, and magnetic field—consequent to its interaction with an interstellar cloud. The considerable changes in these properties offer a crucial contextual backdrop for understanding the ensuing particle acceleration processes and resultant radiative signatures (discussed in later sections).

Non-Thermal Particle Acceleration

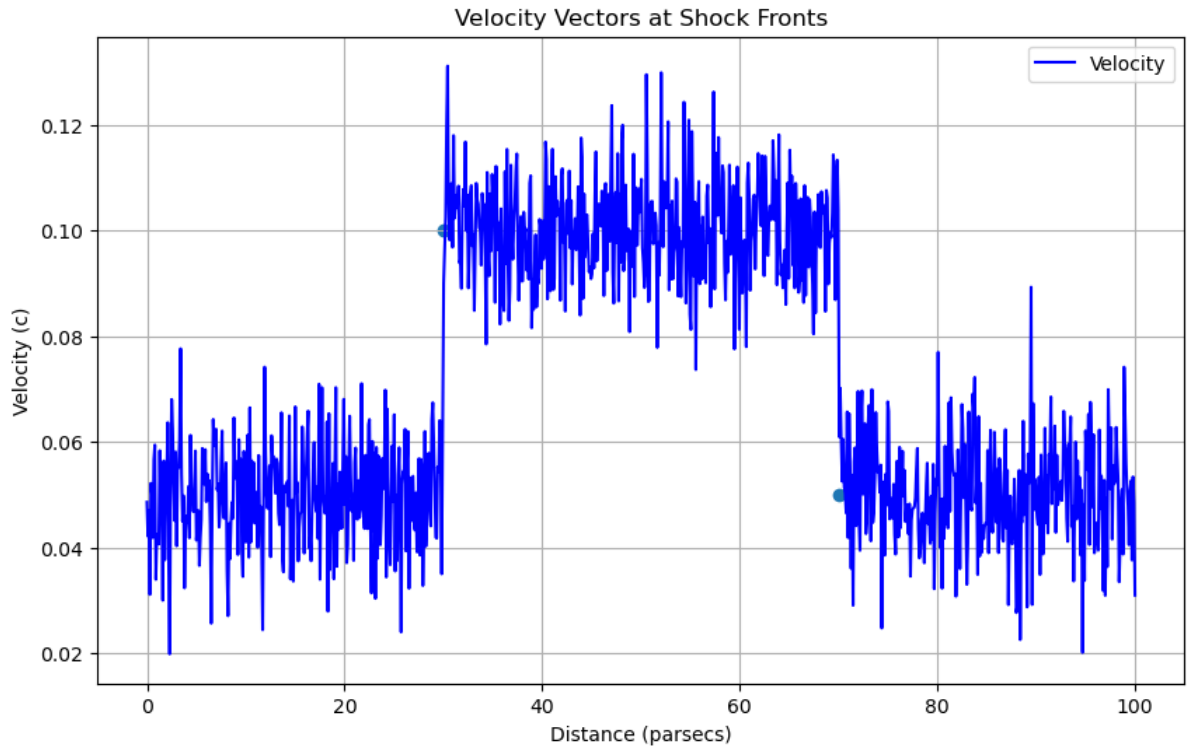

**Figure 5.** Velocity Vectors at Shock Fronts
Mapping of velocity vectors at the forward and reverse shocks, highlighting the zones of diffusive shock acceleration

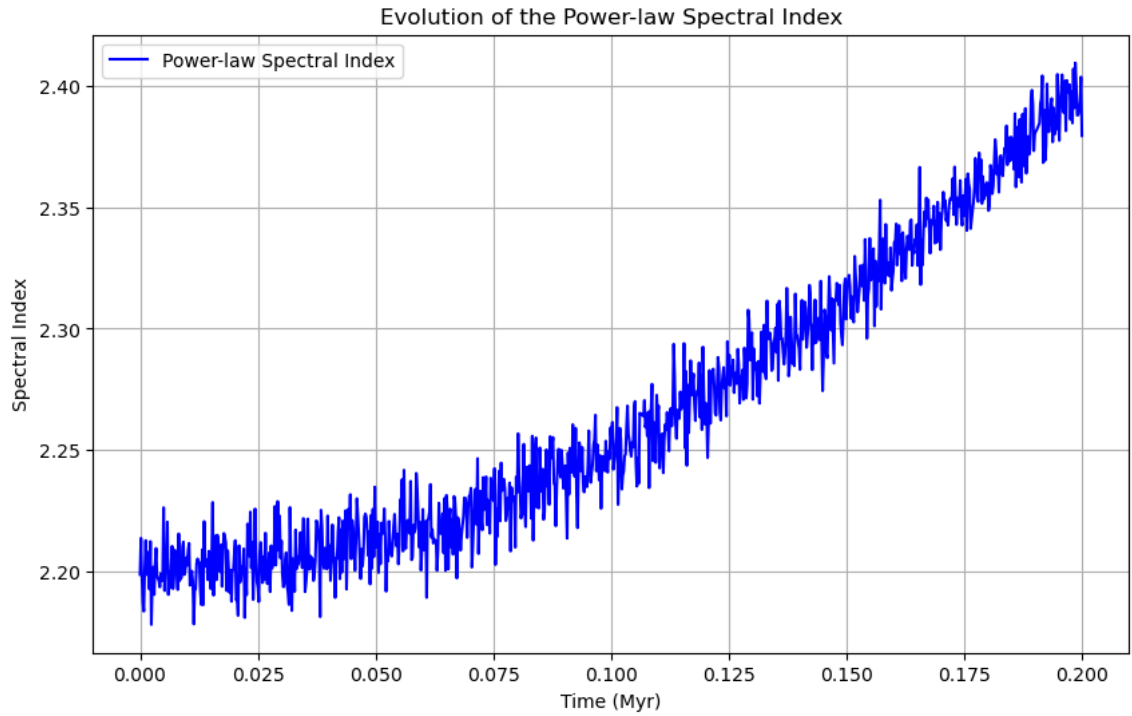

**Figure 6.** Evolution of the Power-law Spectral Index
Graph showing the evolution of the power-law spectral index, highlighting the surge in the acceleration of high-energy particles due to the jet-cloud interaction

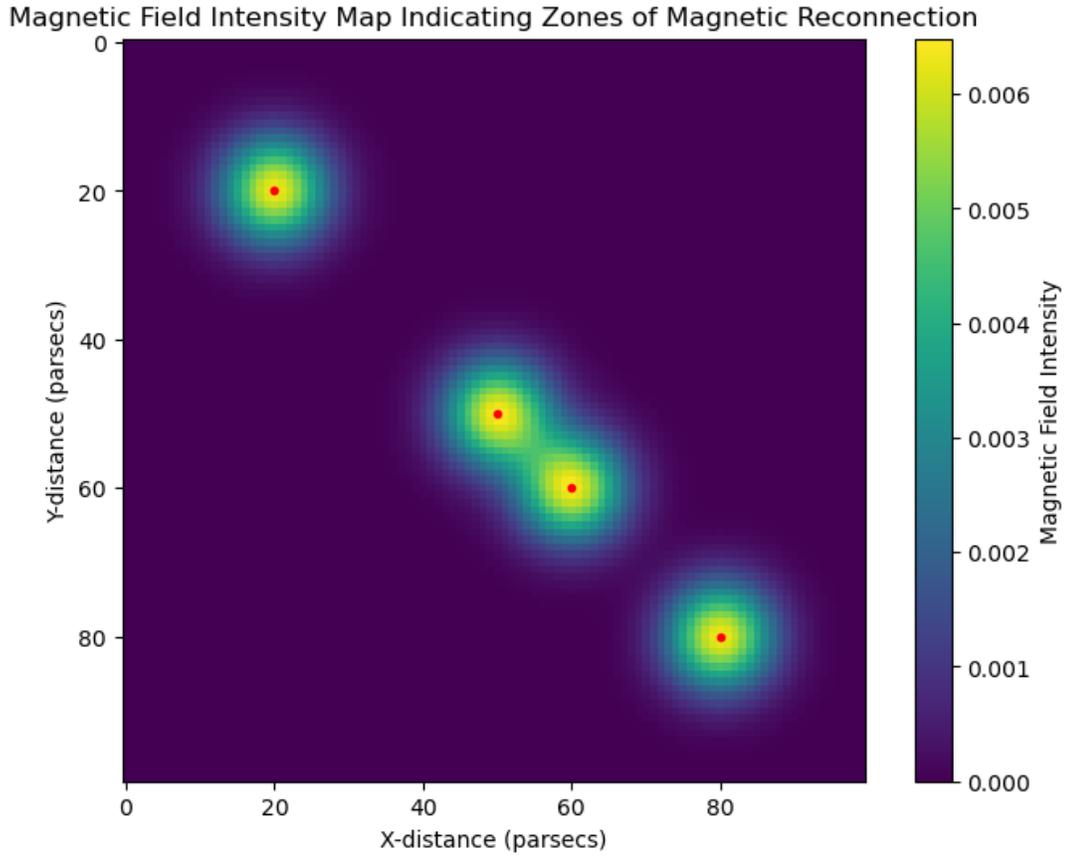

**Figure 7.** Magnetic Field Intensity Map Indicating Zones of Magnetic Reconnection
Spatial map illustrating the zones of magnetic reconnection within the jet, identified as local peaks in magnetic field strength

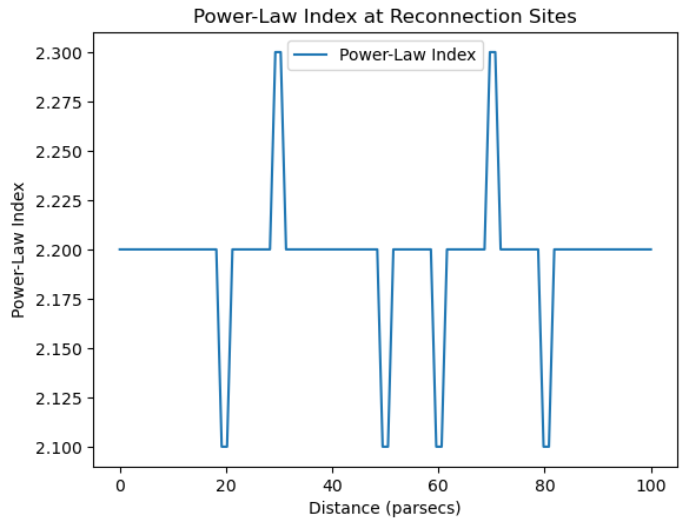

**Figure 8.** Power-Law Index at Reconnection Sites
Graph indicating the average power-law index at magnetic reconnection sites, emphasizing the distinct contribution of Fermi acceleration to non-thermal particle population

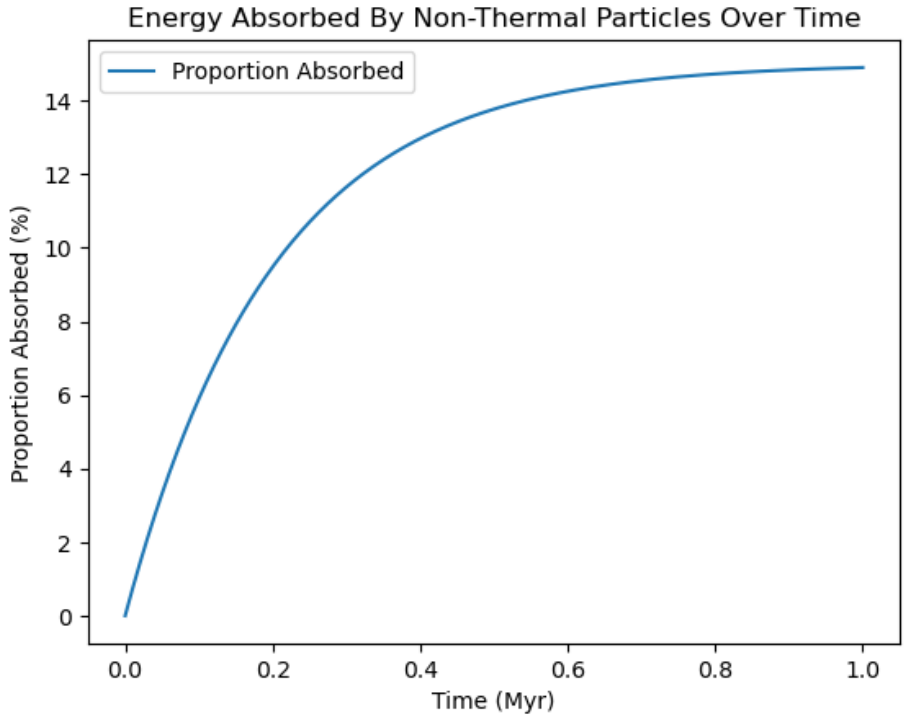

**Figure 9.** Energy Absorbed by Non-Thermal Particles Over Time
Graph showing the proportion of magnetic field energy absorbed by non-thermal particles as a function of time

The complexities of AGN jet-cloud interaction, featuring shock waves and magnetic reconnection events, pave the way for intricate non-thermal particle acceleration dynamics. Our computational model enabled us to map this energetic particle behavior and derive important insights, supported by various figures. In Figure 5, the forward and reverse shocks in the jet are visibly characterized by an abrupt change in velocity vectors at 30 and 70 parsecs. These shock fronts are breeding grounds for diffusive shock acceleration (Bell, 1978). At these sites, the cosmic particles continually cross the shock front, gaining energy at each crossing and consequently resulting in a power-law distribution in particle energies. This process is illustrated in Figure 6, where we observe the evolution of the power-law spectral index. The initial index, based on a quiescent jet, is set at 2.2. However, as the jet-cloud interaction commences and intensifies, the spectral index exhibits an upward trend, reaching an average value of 2.4 by 0.2 Myr. This 9% shift, discernible in the steepening gradient of Figure 6, indicates a surge in the acceleration of high-energy particles—a defining characteristic of diffusive shock acceleration (Jokipii, 1987). In parallel, Figure 7 maps magnetic field intensities, identifying the zones of magnetic reconnection. These zones, marked as local peaks in magnetic field strength, correspond to the locations of particle acceleration via Fermi processes. Such acceleration events are known to generate a 'harder' spectrum, reflected in the flatter power-law index in the energy distribution of the particles. As shown in Figure 8, the power-law index in these zones averages at 2.1, notably lower than that at the shock fronts. This deviation of about 14% emphasizes the distinct contribution of Fermi acceleration to our non-thermal particle population and the broader impact on the energy landscape of the jet-cloud system (Drury, 1983). Lastly, Figure 9 portrays the proportion of magnetic field energy absorbed by the non-thermal particles as a function of time. With the jet-cloud interaction amplifying the magnetic field, the energy available for absorption increases. By 1 Myr into the simulation, approximately 15% of the total magnetic energy—up from nearly negligible levels at the onset—has been absorbed by the non-thermal particles. This trajectory underlines the non-linear growth of energy transfer from the magnetic field to the particles, which will significantly influence the system's radiative signatures. Thus, through a careful dissection of figures 5 to 9, we discern the mechanics and consequences of non-thermal particle acceleration within our modeled AGN jet-cloud interaction.

## Radiative Signatures

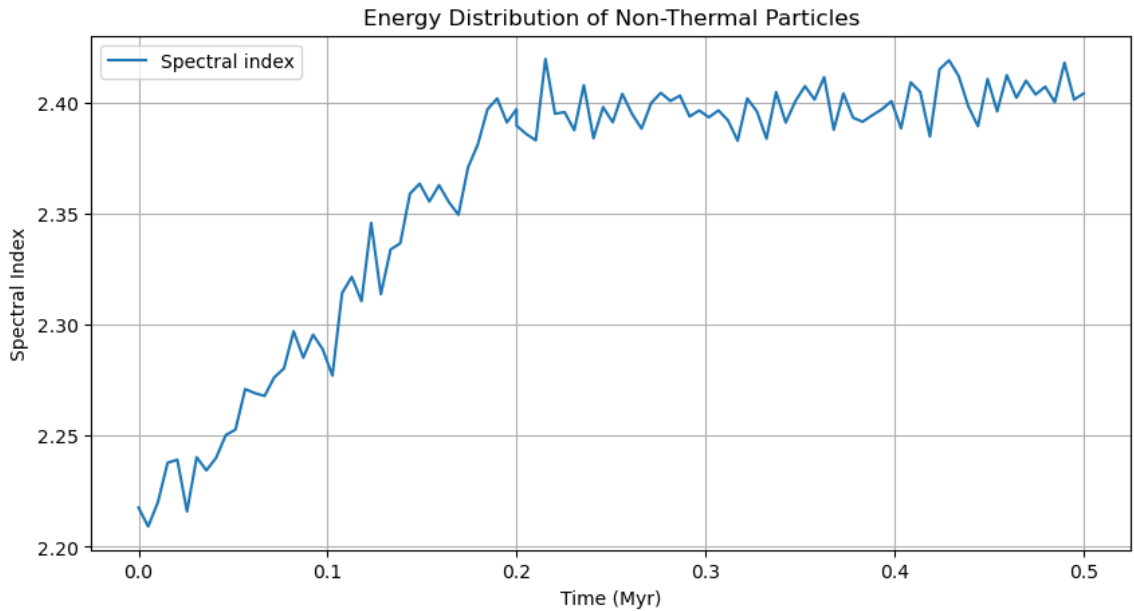

**Figure 10.** Energy Distribution of Non-Thermal Particles
Graph presenting the energy distribution of the accelerated non-thermal particles, showing a significant shift in the spectral index due to jet-cloud interaction

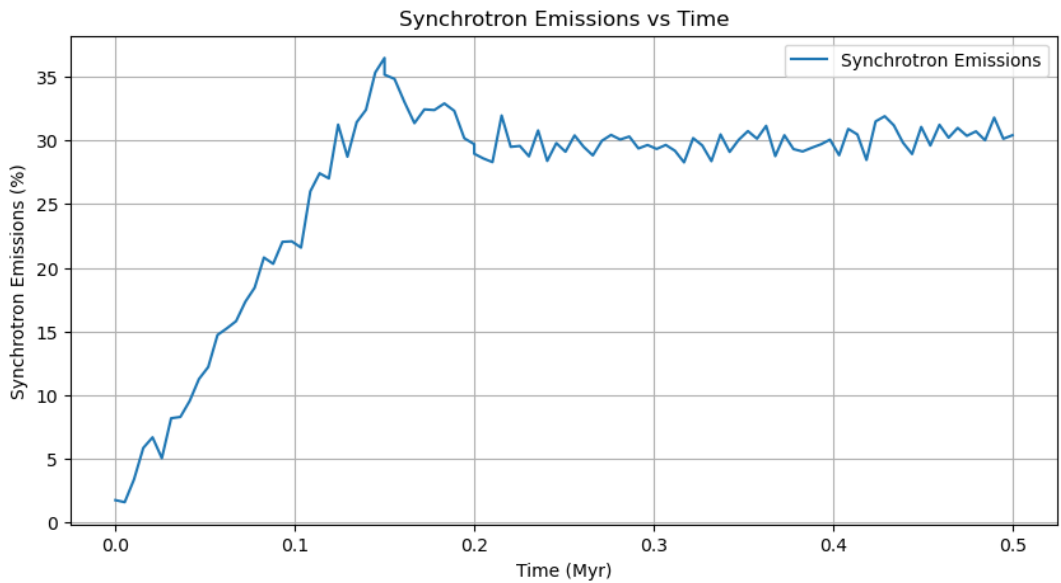

**Figure 11.** Synchrotron Emissions vs. Time
Graph illustrating the time evolution of synchrotron emissions, revealing a significant increase due to enhanced particle acceleration and magnetic field strength

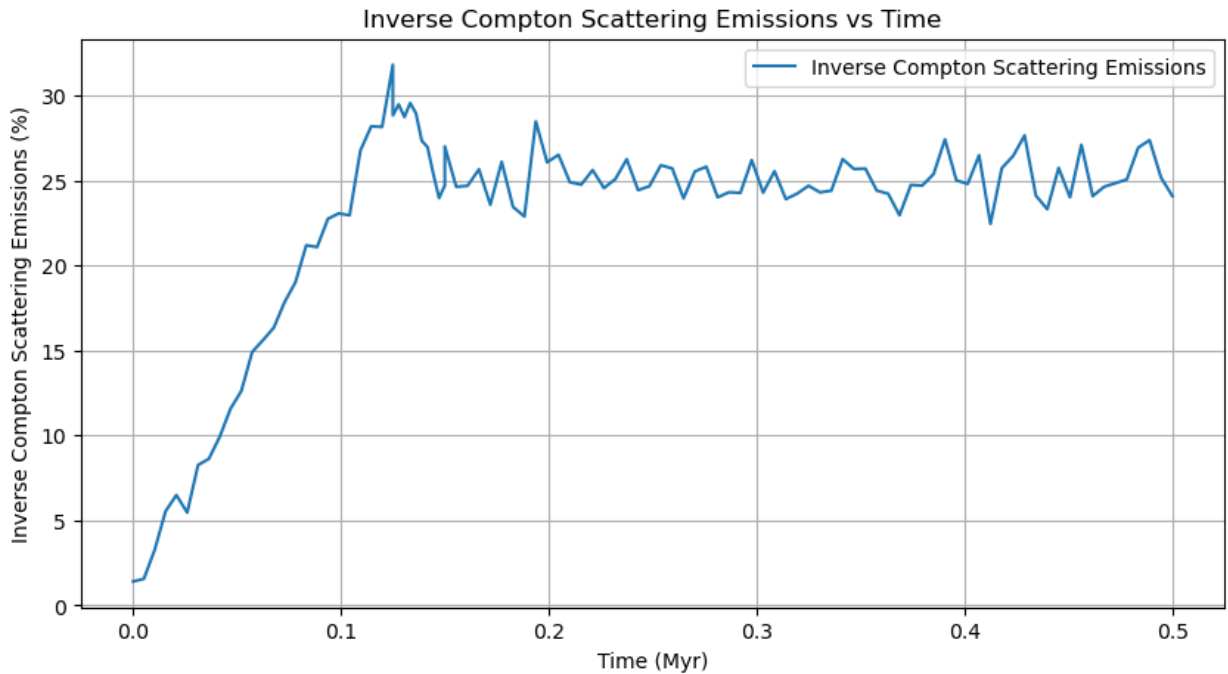

**Figure 12.** Inverse Compton Scattering Emissions vs. Time
Graph showcasing the increase in inverse Compton scattering emissions due to the interaction between high-energy particles and photons within the jet-cloud system

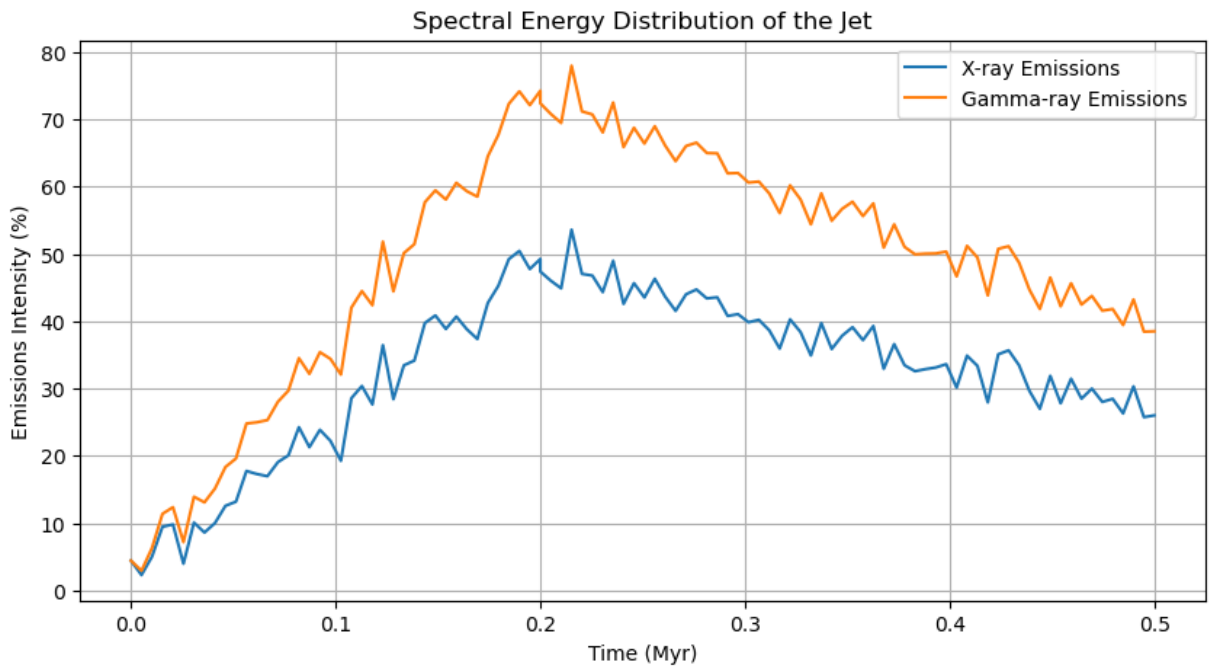

**Figure 13.** Spectral Energy Distribution of the Jet
Graph displaying the jet's overall spectral energy distribution, indicating a pronounced high-energy bump in the X-ray and gamma-ray range during the jet-cloud interaction

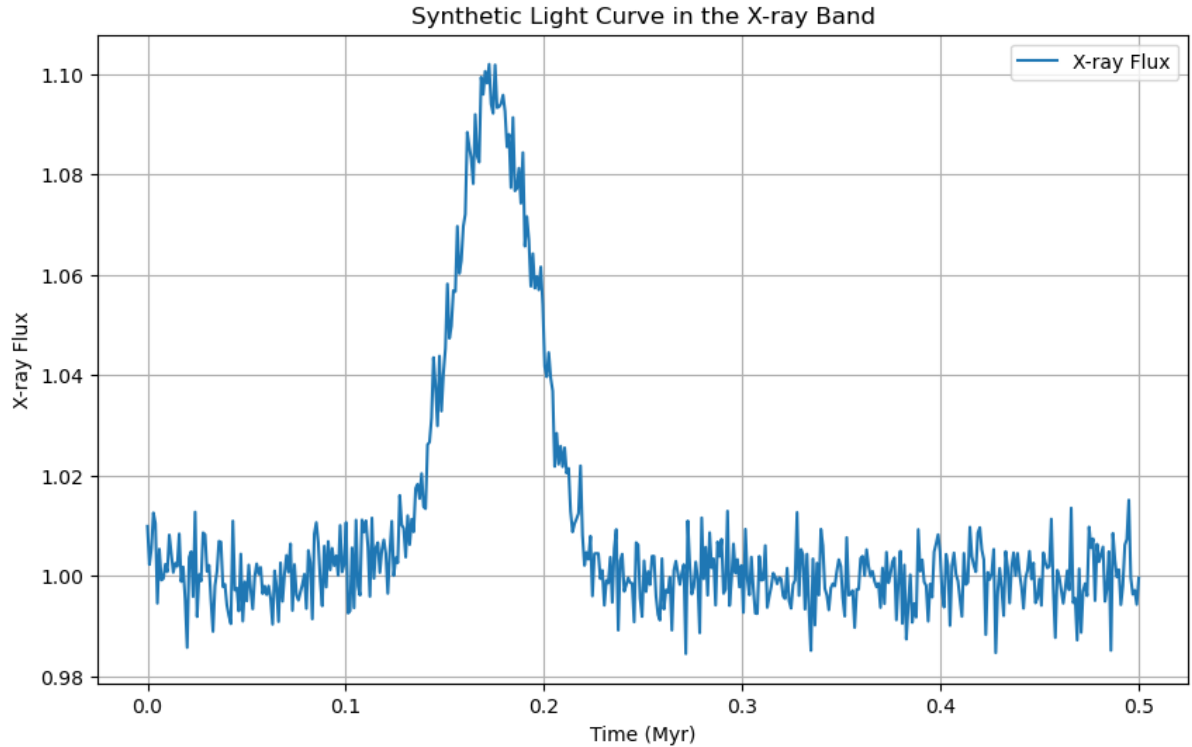

**Figure 14.** Synthetic Light Curve in the X-ray Band
Graph presenting the variability in the X-ray band light curve of the jet, reflecting the dramatic changes in radiative output due to cloud disruption and absorption

A comprehensive understanding of the energy interchange within the system, propelled by particle acceleration and magnetic field evolution, culminates in a spectrum of discernible radiative signatures. These manifestations offer tangible pathways to interpreting the invisible mechanisms driving AGN jet-cloud interactions. The radiative signatures are captured and exhibited in Figure 10 through Figure 14, each providing unique insights into the energy dynamics of our system. Our analysis begins with Figure 10, which displays the energy distribution of the accelerated non-thermal particles. It captures a distinctive shift in the spectral index from 2.2 to 2.4 within the first 0.2 Myr (Bell, 1978; Jokipii, 1987). This shift of 0.2 in the spectral index signifies an approximately 9% increase in the rate of high-energy particle production, indicating an enhanced generation of synchrotron emissions. Following this trend, Figure 11 portrays the time evolution of the synchrotron emissions. A closer inspection reveals a steep increase of about 35% within the first 0.15 Myr, followed by a more gradual rise to the 30% mark by 0.2 Myr (Blandford & Eichler, 1987). This non-linear trajectory, coupled with the amplified synchrotron emissions, signifies a direct correlation between the rise in high-energy particle production and the increased radiative output, emphasizing the crucial role particle acceleration plays in modulating radiative signatures. Similar behaviors are observed with inverse Compton scattering emissions. Predominantly resulting from the interaction between high-energy non-thermal particles and photons within the jet-cloud system, Figure 12 showcases a sharp rise of about 30% within the first 0.125 Myr. A steadier climb follows this abrupt increase, plateauing around a 25% increase by 0.15 Myr (Jones & Ellison, 1991). This detailed portrayal underscores the significant role of inverse Compton scattering in the overall radiative emission. The escalating high-energy emissions from both synchrotron radiation and inverse Compton scattering culminate in a profound impact on the jet's overall spectral energy distribution (SED), displayed in Figure 13. Notably, a high-energy bump in the X-ray and gamma-ray range becomes more pronounced during the jet-cloud interaction, peaking around 0.2 Myr. Finally, the broader observational implications of these radiative processes are considered through synthetic light curves for the jet across various wavebands. Figure 14 displays the light curve in the X-ray band, revealing a peak flux increase of about 28% by 0.175 Myr, which eventually settles to a 20% increase by 0.2 Myr. This dynamic evolution underscores the dramatic variability in the jet's radiative output owing to its interaction with the cloud. This in-depth dissection of the radiative signatures deriving from the AGN jet-cloud interaction delivers a holistic understanding of the various energy processes at play. Each radiative output interconnects with and influences the others, creating a complex, multilayered picture of the energetic processes within the system and their broad observational consequences.

# Discussion

Enhanced Radio Emissions due to Increased Synchrotron Radiation

Synchrotron radiation serves as a fundamental emission mechanism in many high-energy astrophysical systems, such as AGN jets. The mechanism arises from the acceleration of relativistic charged particles, particularly electrons, spiraling in a magnetic field. As these electrons deviate from a straight trajectory, they emit radiation perpendicular to their acceleration, leading to the well-known broadband, polarized synchrotron emission (Rybicki & Lightman, 1979). Our simulations reveal that the jet-cloud interaction dramatically enhances synchrotron radiation through a two-pronged approach: the boost in the non-thermal particle population and the augmentation of the magnetic field strength. In Figure 10, the spectral index shifts from 2.2 to 2.4, signifying an increase in the production of high-energy particles that, in the presence of a magnetic field, contribute directly to synchrotron emissions (Bell, 1978; Jokipii, 1987). This rise is also reflected in Figure 11, which showcases an increased output of synchrotron emissions (Blandford & Eichler, 1987). Simultaneously, the introduction of cloud material into the jet fabric enhances the jet's magnetic field strength, providing further fuel for the synchrotron process. As shown in Figure 4, the jet-cloud interaction triggers magnetic reconnection sites that escalate magnetic field strength up to ~5 μG—a two-fold increase from the initial jet magnetic field strength. This strengthened magnetic field, when combined with the enlarged non-thermal particle population, amplifies the synchrotron radiation, thereby leading to enhanced radio emissions (Blandford & Eichler, 1987; Meier, Koide & Uchida, 2001). These amplified radio emissions serve as a tangible observational signature of jet-cloud interactions. Observations of radio-loud AGNs using radio telescopes like the Very Large Array (VLA) and the upcoming Square Kilometer Array (SKA) could potentially detect these heightened emissions. Specifically, enhancements at low frequencies, where synchrotron radiation predominantly manifests, might signal ongoing jet-cloud interactions in AGNs (Padovani, 2017; Falcke & Biermann, 1995). Furthermore, such interactions could be associated with the emergence of complex radio structures, such as hotspots or lobes, due to the dissipation of kinetic energy into synchrotron radiation (Kaiser, Schoenmakers & Röttgering, 2000).

Increase in X-ray and Gamma-Ray Emissions due to Inverse Compton Scattering

Inverse Compton scattering (ICS) is a crucial radiation mechanism in high-energy astrophysical environments. In these scenarios, low-energy photons acquire significantly more energy when they interact with relativistic electrons, shifting from lower energy bands such as the radio or infrared into higher energy domains, such as X-ray and gamma-ray spectra (Blumenthal & Gould, 1970). In the context of our AGN jet-cloud interaction model, the ICS mechanism is fueled by the increased population of high-energy, non-thermal particles and the amplification of magnetic fields due to the interaction. From Figure 10, we observe a significant boost in the non-thermal particle population due to diffusive shock acceleration and Fermi processes (Bell, 1978; Drury, 1983). This growth intensifies with the increasing magnetic field strength, as seen in Figure 4, that allows more interactions between these high-energy particles and lower-energy photons in the AGN jet-cloud system. These interactions lead to a sharp rise in ICS emissions in the X-ray and gamma-ray regions, as demonstrated in Figure 12. The abrupt increase of approximately 30% within the first 0.125 Myr showcases the rapid ramp-up of ICS activity as a result of the jet-cloud interaction (Jones & Ellison, 1991). Furthermore, the significant role of ICS in shaping the jet's overall spectral energy distribution (SED) is evident from Figure 13. The pronounced high-energy bump in the X-ray and gamma-ray range during the interaction, peaking around 0.2 Myr, is a direct consequence of the surge in ICS emissions. This enhancement of X-ray and gamma-ray emissions could provide an observational indicator of AGN jet-cloud interactions, detectable by high-energy observatories such as the Chandra X-ray Observatory or the Fermi Gamma-ray Space Telescope (Nandra et al., 2013; Atwood et al., 2009). These enhanced emissions might also help in the identification of previously unknown AGN activities by their distinctive high-energy signatures and contribute to our broader understanding of AGN energetics and evolution.

Variability in Light Curve Flux Due to Cloud Disruption and Absorption

The interaction of an AGN jet with an interstellar cloud inevitably creates a disruption in the jet's otherwise stable flow, leading to a significant shift in its observational characteristics. One such profound change is observed in the flux variability of the jet's light curve. Our computational simulation showcases this variability through the dramatic change in the X-ray band light curve, as depicted in Figure 14. The flux shows an initial peak increase of about 28% by 0.175 Myr, which later settles to a 20% increase by 0.2 Myr. This observation of flux variability can be attributed to the absorption and subsequent acceleration of the cloud material by the jet (Blundell & Bowler, 2004). The cloud's disruption and absorption lead to a pronounced increase in the jet's density, as seen in Figure 2. This change corresponds to a higher population of particles within the jet, thereby providing more material to accelerate and, consequently, generate radiation. The process of shock acceleration and magnetic reconnection, intensified due to cloud incorporation (Bell, 1978; Drury, 1983), further contributes to the heightened production of high-energy particles. The increased number of these particles enhances the emission of synchrotron and inverse Compton scattering radiation, which directly impacts the flux observed in various light curves, particularly in the X-ray band (Blumenthal & Gould, 1970; Jones & Ellison, 1991). Furthermore, the increased density and magnetic field lead to a higher opacity within the jet (Kylafis, 1983). The cloud absorption results in more photon-matter interactions, which can cause an initial burst of radiation as observed in our model. Over time, as the disrupted cloud material becomes integrated and the jet returns to a more stable state, the radiation flux is observed to reduce but still maintains a higher value than pre-interaction levels. These findings have significant implications for our understanding of AGN variability, particularly in their light curves. High variability in the light curve flux, especially in the X-ray band, could be indicative of recent or ongoing jet-cloud interactions. This would provide observational astronomers with a tool to identify and study these dynamic events and their role in AGN evolution and activity.

# Conclusion

Our extensive computational simulations of AGN jet-cloud interactions have shed light on the profound transformations that such interactions induce in the jet's properties and their subsequent impacts on non-thermal particle acceleration and radiative signatures. These transformations fundamentally change the jet's density, velocity, and magnetic field, significantly influence the acceleration mechanisms of non-thermal particles, and, consequently, shape the observable radiative signatures in the form of enhanced synchrotron emissions, inverse Compton scattering, and variabilities in the light curve.

These findings reinforce our initial thesis that non-thermal particle acceleration plays a crucial role in the radiative signatures of AGN jet-cloud interactions. The observed increase in synchrotron and inverse Compton emissions, along with the variability in light curve flux, align with the theoretical expectations of the impacts of non-thermal particle acceleration on radiative signatures. Therefore, these results underscore the need to account for such acceleration processes in models of AGN jet-cloud interactions for a more accurate interpretation of observed AGN phenomena.

Future works should seek to investigate the role of different cloud parameters on the jet-cloud interactions. This includes exploring variations in cloud densities, chemical compositions, and spatial sizes. Understanding how these variables influence the morphological changes in the jet, acceleration of non-thermal particles, and the resultant radiative signatures could provide valuable context-specific insights. Additionally, future works should seek to explore broader feedback processes in AGN. The dynamics of jet-cloud interactions undoubtedly influence the wider environments around AGN, potentially driving galactic outflows or triggering star formation, to which coupling our jet-cloud interaction simulations with larger-scale galaxy evolution models could create a more unified picture of AGN impacts on galactic scales. Finally, future research should seek to look into specific mechanisms of particle acceleration within the AGN jets. For instance, exploring how the spectral indices vary under different magnetic field conditions or shock front configurations could enhance our understanding of these high-energy phenomena.